\documentclass[aps,pre,amsmath,amssymb,superscriptaddress,twocolumn]{revtex4-1}
\usepackage[T1]{fontenc}
\pdfoutput=1
\usepackage[dvipsnames,rgb,dvips]{xcolor}
\usepackage{graphicx}
\usepackage{float}
\usepackage{overpic}
\usepackage{amsmath,amssymb,amsfonts,MnSymbol}
\usepackage{hyperref}
\makeatletter
\def\amsbb{\use@mathgroup \M@U \symAMSb}
\makeatother
\usepackage{mathrsfs}
\usepackage{overpic}
\newcommand\nn{\nonumber}

\newcommand{\ee}{\ensuremath{\text{e}}}
\newcommand{\ed}{\ensuremath{\text{d}}}
\newcommand{\Ku}{\ensuremath{\text{Ku}}}
\newcommand{\St}{\ensuremath{\text{St}}}
\newcommand{\tauc}{\ensuremath{\tau_\text{c}}}
\newcommand{\taup}{\ensuremath{\tau_\text{p}}}
\newcommand{\taua}{\ensuremath{\tau_\text{a}}}

\newcommand{\tauk}{\ensuremath{\tau_\text{K}}}

\graphicspath{{figures/}}
\begin{document}
\title{Heavy particles in a persistent random flow with traps}
\author{J. Meibohm}
\affiliation{Department of Physics, Gothenburg University, SE-41296 Gothenburg, Sweden}
\author{B. Mehlig}
\affiliation{Department of Physics, Gothenburg University, SE-41296 Gothenburg, Sweden}

\begin{abstract}
We study a one-dimensional model for heavy particles in a compressible fluid. The fluid-velocity field is modelled by a persistent Gaussian random function, and the particles are assumed to be weakly inertial.
Since one-dimensional fluid-velocity fields are always compressible, the model exhibits spatial trapping regions where particles tend to accumulate. We determine the statistics of fluid-velocity gradients in the vicinity of these traps and show how this allows to determine the spatial Lyapunov exponent and the rate of caustic formation. 
We compare our analytical results with numerical simulations of the model and explore the limits of validity of the theory. Finally, we discuss implications for higher-dimensional systems.
\end{abstract} 
\maketitle

\section{Introduction}

Turbulent fluids in Nature and Technology often contain small particles whose density is larger than that of the carrier fluid. Examples are sprays, water droplets in turbulent clouds, and dust in the exhaust of combustion engines. It is a common observation that these particles do not distribute homogeneously in space, but instead form regions of high and low concentration on small scales. This effect is referred to as {\em spatial clustering} \cite{Bec03,Wil03,Gus16}, and it  is a consequence of the interaction of the particles with the turbulent fluid. Hence, any attempt to describe these systems must take into account the turbulent fluctuations, a challenging problem by itself.

Recently, much progress has been made in explaining and quantifying spatial clustering of heavy particles in turbulence, using statistical models. See Ref.~\cite{Gus16} for a recent review. Such models approximate the turbulent fluid-velocity fluctuations below the smallest scales of turbulence, called the dissipative range, in terms of random velocity fields with prescribed statistics. Clearly, these random velocity fields lack the complexity of real turbulence. Remarkably, however, such statistical models often reproduce the dynamics of the immersed particles  and the statistical properties of their spatial distribution qualitatively, and in some cases even quantitatively. Statistical models allow for much simpler numerical simulations and even admit analytical treatment in limiting cases \cite{Wil03, Gus16, Fal07}. The models typically depend on two parameters, the Kubo number $\Ku$, which is a measure of the persistence of the flow, and the Stokes number $\St$, which measures the importance of particle inertia in the fluid-particle system.

Spatial clustering is often characterised in terms of the spatial Lyapunov exponents $\lambda_i$, $i=1,\ldots,d$ \cite{Som93,Gus16}. These exponents measure the rates at which particles approach or diverge, in the long-time limit. Because heavy particles are not constrained to follow the flow, their dynamics takes place in $2d$-dimensional phase space, while their spatial concentration is measured in the $d$-dimensional coordinate space. As a consequence, the spatial distribution of the particles exhibits singularities, so-called caustics, which contribute to the spatial clustering of the particles \cite{Fal02,Wil05,Wil06,Fal07}. One therefore identifies the rate of formation of these singularities, the rate of caustic formation $\mathcal{J}$, as a second important observable.

A well-studied limiting case of the statistical model is the so-called white-noise limit, describing the dynamics of very heavy particles in a rapidly fluctuating fluid-velocity field. In terms of the dimensionless parameters, this limit corresponds to $\Ku\to0$ and $\St\to\infty$ so that $\Ku^2\St$ remains constant \cite{Gus16}. An important property of white-noise models is that they depend solely on a single parameter, $\varepsilon^2 = f(d,\beta)\, \Ku^2 \St$, where $f$ is a function of the spatial dimension $d$ and the compressibility parameter $\beta$ of the underlying flow field. The parameter $\varepsilon$ takes the role of an inertia parameter in the white-noise model, similar to the Stokes number $\St$ in statistical models at finite values of $\Ku$ \cite{Gus16}. In the white-noise limit in one spatial dimension, $\lambda = \lambda_1$ and $\mathcal{J}$ can be computed analytically as a function of $\varepsilon$ \cite{Wil03}.

In this Paper, we describe a second limiting case of the statistical model, the persistent-flow limit. This limit describes weakly inertial, heavy particles in a very persistent flow, hence the name. The limit is taken by letting $\Ku\to\infty$ and $\St\to0$ such that the product $\Ku\,\St$ stays finite. Similarly to the white-noise limit, the persistent-flow limit turns out to depend on a single inertia parameter $\kappa$, where $\kappa^2 = (d+2) \Ku^2 \St^2$ \cite{Gus16}.

Here, we confine ourselves to the one-dimensional version of the model, $d=1$, and show that $\lambda$ and $\mathcal{J}$ can be computed analytically in the persistent-flow limit. In one spatial dimension, the random fluid-velocity field is compressible. Because of the high persistence of the flow, compressibility leads to long-lived trapping regions, regions in space where the particles accumulate. We show that the statistics of fluid-velocity gradients in the trapping regions are the crucial ingredient to solving the one-dimensional model.

The Paper is organised as follows: In Sec.~\ref{sec:prob} we introduce the model for heavy particles in turbulence and motivate the persistent-flow limit by a detailed analysis of the relevant time scales. Section~\ref{sec:res} contains the statement and the derivation of the analytical results for the spatial Lyapunov exponent $\lambda$ and the rate of caustic formation $\mathcal{J}$. We compare these results to those obtained in the  white-noise limit,  and to  results of numerical simulations of the statistical model. We draw conclusions and discuss implications for higher-dimensional models in Sec.~\ref{sec:conc}. The Appendix contains an alternative derivation of the main formulae derived in Sec.~\ref{sec:res}.

\section{Problem formulation}\label{sec:prob}
We study a one-dimensional model for the motion of heavy particles in the dissipative range of turbulence, approximating the viscous force on the particle by Stokes drag. 
The equations of motion for an underdamped small and heavy particle in a viscous fluid are given by
\begin{align}\label{eq:eom}
	\dot x(t) = v(t)\,,	\qquad	\dot v(t) = \tau_\text{p}^{-1}\left[ -v(t) + u(x(t),t) \right]	\,,
\end{align}
where $\tau_\text{p}=\gamma^{-1}$ is the particle relaxation time, the typical time it takes the particle to relax to the fluid velocity. We model the turbulent fluid velocity $u(x,t)$ by a one-dimensional Gaussian random velocity field with zero mean and correlation \cite{Gus16}
\begin{align}\label{eq:corr}
	\langle u(x,t) u(x',t') \rangle = u_0^2 \big[1\!-\!(x\!-\!x')^2/\eta^2\big]	\ee^{-(x-x')^2/\eta -|t-t'|/\tauc}\,.
\end{align}
In this formulation, the flow field is a function of its root mean square velocity $u_0$ as well as its correlation length and fluid correlation time, $\eta$ and $\tauc$, respectively.  Fields in one spatial dimension are always potential, therefore the velocity field $u(x,t)$ is compressible, as mentioned in the Introduction. Models for particles in three-dimensional turbulence, by contrast, usually assume that the flow is incompressible.
 Further, approximating $u(x,t)$ by a Gaussian random field  neglects several characteristic aspects of turbulence such as intermittency and broken time-reversal invariance. It has, however, been shown that the highly-simplified one-dimensional model gives important insights into the dynamics of particles in the dissipative range of turbulence (see \cite{Gus16} and references therein).
\subsection{Relevant time scales}
The random velocity field $u(x,t)$ is characterised by two different time scales. In addition to the correlation time $\tauc$, the advection time scale $\taua = \eta/u_0$ is the typical time it takes an advected (tracer) particle to travel one correlation length $\eta$. From the three time scales $\tauc$, $\taua$ and $\taup$, one can define three dimensionless numbers \cite{Gus16}
\begin{align}\label{eq:dlnum}
	\Ku = \tauc \taua^{-1}\,, \quad \St = \taup \tauc^{-1}	\, \quad \text{and}\quad \kappa = \sqrt{3}\,\taup \taua^{-1}\,.
\end{align}
The Kubo number, $\Ku$, measures the persistence of the flow and $\St$ is the Stokes number, a measure for particle inertia in the system. The third quantity in expression \eqref{eq:dlnum}, $\kappa$, compares the particle relaxation time to the advection time and is thus another measure of particle inertia.

In case there is a time-scale separation between $\taua$ and $\tauc$, only one of the two inertia parameters, either $\St$ or $\kappa$, needs to be considered. In the white-noise limit \cite{Wil03,Wil05,Wil06,Der07,Gus16,Pum16}, for instance, the correlation time $\tauc$ is the \textit{smallest} of all time scales. More precisely, this limit is approached if $\taup\gg\taua\gg\tauc$ so that the parameter $\varepsilon\propto\sqrt{\taup\tauc}\taua^{-1} = {\rm Ku}\sqrt{ {\rm St}}$ stays finite. In other words, the white-noise limit of Eqs.~\eqref{eq:eom} describes very heavy particles in a rapidly fluctuating flow field. The fluid-velocity gradient becomes a white-noise signal with zero mean and correlation strength $\propto\varepsilon^2$. Thus, Eqs.~\eqref{eq:eom} constitute a generalised diffusion process and one can use Fokker-Planck equations to solve the model exactly in one dimension, $d=1$ \cite{Wil03,Gus16}. The single relevant dimensionless inertial parameter in the white-noise model is $\varepsilon$, which is related to the diffusion constant $D$ of the generalised diffusion by $D\taup^{-1}\propto\varepsilon^2\propto \Ku^2 \St$.

In the persistent-flow limit studied here, by contrast, the correlation time $\tauc$ is the \textit{largest} of the three time scales, i.e., $\tauc\gg \taup,\taua$. In terms of dimensionless parameters this limit corresponds to $\Ku \gg 1$ and $\St \ll 1$. The ratio between the remaining time scales is thus given by $\kappa \propto \taup/\taua$. Hence, the particle motion and the motion of advected particles take place on time scales much smaller than the correlation time of the fluid. The inertia parameter $\kappa$ measures how much the particles detach from the flow, compared to the motion of advected tracers.f

The advection time scale $\taua$ in the statistical model is analogous to the Kolmogorov time $\tauk$ in homogeneous isotropic turbulence, defined as $\tau_{\rm K} = 1/\sqrt{{\rm tr}\,\langle\mathbb{A}\mathbb{A}^{\sf T}\rangle}$.
Here $\mathbb{A}$ is the matrix of fluid-velocity gradients, and $\langle \cdots\rangle$ is an average along
fluid-element paths. Since the Kolmogorov time is the only available microscopic time scale in turbulence, the analogue of $\kappa$ (constructed from $\tauk$ and $\taup$), is commonly used as the inertia parameter in DNS and experiments on the dynamics of heavy particles in turbulence \cite{Per12, Gus16}.

\subsection{One-dimensional model}
The observables $\lambda$ and $\mathcal{J}$ are most conveniently studied in terms of the relative dynamics of an ensemble of particle pairs, instead of the single-particle dynamics given in Eqs.~\eqref{eq:eom}. The motion of a pair of particles is described by the particle separation $X = x^{(1)} - x^{(2)}$ and relative velocity $V = v^{(1)} -v^{(2)}$, as well as their midpoint position $\bar x = (x^{(1)}+x^{(2)})/2$ and mean velocity $\bar v = (v^{(1)}+v^{(2)})/2$. At separation much smaller than $\eta$, $X\ll\eta$, a particle pair experiences an essentially linear velocity field. We therefore linearise the equations of motion \eqref{eq:eom} with respect to $X$. To this end, we introduce the mean flow $\bar u (\bar x,t) \sim \left[u(x^{(1)},t)+u(x^{(2)},t)\right]/2$ and the fluid-velocity gradient, $A(\bar x,t) \sim \left[u(x^{(1)},t)-u(x^{(2)},t)\right]/X$. A particle pair at small separation then obeys the linearised equations of motion
\begin{subequations}\label{eq:eomX}
    \begin{alignat}{4}
     	& \dot X &&= V\,,	\qquad &&\dot V &&= \taup^{-1}\left[A(\bar x,t)X - V \right]	\,,	\label{eq:eomXa}\\
	& \dot{\bar x} &&= \bar v\,, \qquad &&\dot {\bar v} &&= \taup^{-1}\left[\bar u(\bar x,t)-\bar v\right]	\,, \label{eq:eomXb}
    \end{alignat}
\end{subequations}
where we have omitted the time dependence of $(X, V, \bar x, \bar v)$ for convenience of notation. The particle-velocity gradient $Z = V / X$, has the equation of motion \cite{Fal02, Wil03,Gus16}
\begin{align}\label{eq:eomZ}	
	\dot Z = \taup^{-1}\left[A(\bar x,t) - Z\right] - Z^2\,.
\end{align}
The two equations for $(\bar x, \bar v)$, Eqs.~\eqref{eq:eomXb}, resemble a closed system that is equivalent to the single-particle dynamics \eqref{eq:eom}. Equations~\eqref{eq:eomXa} and \eqref{eq:eomZ}, on the other hand, are not closed as they depend on the variable $\bar x$ which enters in the argument of the fluid-velocity gradient.

As both advected tracers and heavy particles move much faster than the fluid changes, $\taua,\taup\ll\tauc$, and the fluid-velocity field $u(x,t)$ is compressible, there exist long-lived, stable fixed points of Eqs.~\eqref{eq:eom}, and thus Eqs.~\eqref{eq:eomXb}, where the particles accumulate. These trapping regions are characterised by negative fluid-velocity gradients $A(x,t)<0$, and small fluid velocity $u(x,t) \ll u_0$. Trapping regions appear and disappear randomly with mean rate proportional to $\tauc^{-1}$ \cite{Meh09}. When a trapping region vanishes, the trapped particles are driven to a neighbouring trapping region, at a typical distance $\sim \eta$ away.

If $\taup\ll\taua$ we estimate the migration time to the next trapping region by the time it takes for advected tracers to travel one correlation length, $\taua$. If $\taup\gg\taua$, on the other hand, the migration time is of the order of the relaxation time $\taup$ of the particles. We conclude that the migration time is of the order of the maximum of these two time scales, $\max(\taup,\taua)$.

 As trapping regions persist for times $\sim\tauc$, the persistent-flow limit $\tauc\gg\taua,\taup$ ensures that the time particles spend in trapping regions is much larger than the migration time. Hence, for $\Ku\gg1$ and $\St\ll1$ we can neglect the disappearance and appearance of stable zeros of $u(x,t)$ and consider the case of an ensemble of particles in an unbounded and frozen Gaussian random velocity field. 
 
Each trapping region in the frozen velocity field is characterised by a constant fluid-velocity gradient $A_n<0$, where we index the trapping regions by $n$, for reasons that become clear later. Because different trapping regions are typically a fluid correlation length $\eta$ away from each other, their fluid-velocity gradients can be assumed to be independent and identically distributed.
 
To obtain the equation of motion for a particle pair in the $n^\text{th}$-trapping region of the flow, we first dedimensionalise Eqs.~\eqref{eq:eomX} and \eqref{eq:eomZ}, so that they do not explicitly dependent on the time scales. That is, we let $t\to\taup t$, $X\to\eta X$, $V \to \eta \taup^{-1} V$, $A\to \taup^{-1} A$ and $Z\to \taup^{-1}Z$. We obtain the dimensionless equations
\begin{subequations}	\label{eq:eomfinal}
\begin{align} 
	   \dot X &= V\,, \quad \dot V = A_n X -V	\,,	\label{eq:eomXVfinal}\\
	   \dot Z &=  A_n - Z -Z^2\,.	\label{eq:eomZfinal}
\end{align}
\end{subequations}
In the rest of this Paper, all quantities, including the Lyapunov exponent $\lambda$ and the rate of caustic formation $\mathcal{J}$, are written in dimensionless form. 

The probability density of the random fluid-velocity gradient in the $n^\text{th}$-trapping region, $A_n$, is given by the distribution of gradients at the zeros of the Gaussian random function $u(x,t)$, conditioned on that these gradients are negative \cite{Gus13b}. This distribution is obtained by means of the \lq{}Kac-Rice formula\rq{} \cite{Kac48,Ric44}, which allows to compute the density of singular points
of random functions. Using that the gradient function $A(x,t)$ is Gaussian distributed, we obtain for the probability density of $A_n$:
\begin{align}\label{eq:kacrice}
	P(A_n = a) = \begin{cases} \frac{|a|}{\kappa^2} \ee^{-a^2	/(2\kappa^2) }		&	a\leq0\,,	\\0	& \text{otherwise}\,.			\end{cases}
\end{align}
Eq.~(\ref{eq:kacrice}) shows, that the particle-velocity distribution depends only on the single parameter $\kappa$ [Eq.~(\ref{eq:dlnum})] in the persistent limit.
Furthermore it follows from Eq.~\eqref{eq:kacrice} and the independence of the gradients $A_n$  that
\begin{subequations}\label{eq:meanvar}
	\begin{align}
		\langle A_n \rangle &= -\sqrt{\tfrac\pi2}\kappa\,,	\label{eq:mean}\\
		\langle A_n A_m \rangle - \langle A_n \rangle \langle A_m \rangle &= 2 \kappa^2\left(1-\tfrac{\pi}{4}\right) \delta_{n m}	\label{eq:var}\,,
	\end{align}
\end{subequations}
for all $n$ and $m$. Clearly, $\langle A_n \rangle < 0$ is the result of  strong preferential sampling of negative gradients, due to the accumulation of particles in trapping regions. Eq.~(\ref{eq:kacrice}) shows, in fact,  that  $P(A_n >0)=0$. 
\section{Results}\label{sec:res}
Using Eqs.~\eqref{eq:eomfinal} and the distribution of $A_n$ we calculate the spatial Lyapunov exponent $\lambda$ and the rate of caustic formation $\mathcal{J}$. Both quantities can be either obtained directly from the equations of motion for $X$ and $V$, Eqs.~\eqref{eq:eomXVfinal}, or from the steady-state probability density of $Z$, $P_s(Z=z)$. We explain the method using $P_s(Z=z)$ in the following Section and describe the first method in Appendix~\ref{sec:obsdirect}. The general idea in both cases is to first compute the respective quantities conditional on a certain gradient realisation, $A_n=a$, and then average over all realisations using the probability density $P(A_n=a)$ given in Eq.~\eqref{eq:kacrice}.
\subsection{Steady-state distribution of $Z$}
\begin{figure}[t]
	\includegraphics{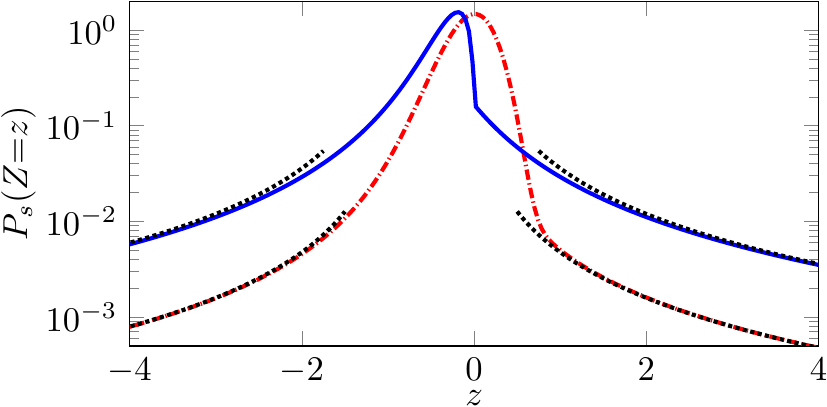}
	\caption{Steady-state distributions for $Z$, $P_s(Z=z)$, as a function of $z$. Shown are the persistent-flow model result \eqref{eq:pz} for $\kappa=0.25$ (solid line) and the corresponding distribution for the white-noise model (dash-dotted line). The dotted lines show the power-law tails $\sim \mathcal{J} z^{-2}$.}\label{fig:zdistr}
\end{figure}
We first turn to the computation of the steady-state distribution of $Z$. The conditional density $\varrho(Z=z|A_n = a)$ obeys the equation
\begin{align}
	\mathcal{F}_{A_n =a} \varrho(Z=z|A_n=a) = \partial_t \varrho(Z=z|A_n=a)\,,
\end{align}
where the operator $\mathcal{F}_{A_n =a}$ is the generator of the Perron-Frobenius operator  \cite{Cvi05} corresponding to the dynamics of $Z$ in Eq.~\eqref{eq:eomZfinal} given the realisation $A_n = a$. The operator reads
\begin{align}
	\mathcal{F}_{A_n =a} = \frac{\ed}{\ed z} (z+z^2 - a)\,.
\end{align}
By definition, the conditional steady-state probability density $\varrho_s(Z=z|A_n=a)$ solves $\partial_t \varrho_s(Z=z|A_n=a)=0$ and is thus annihilated by $\mathcal{F}_{A_n =a}$,
\begin{align}\label{eq:rhozeq}
	\frac{\ed}{\ed z} (z+z^2 - a) \varrho_s(Z=z|A_n=a)=0	\,.
\end{align}
Equation~\eqref{eq:rhozeq} has the solution
\begin{align}\label{eq:zconddist}
	\varrho_s(Z=z|A_n=a) = \begin{cases} \frac{\sqrt{-1-4a}}{2 \pi  \left(z^2+z-a\right)}	& a < -1/4	\\
								\delta(z-z^*)							& a \geq -1/4
					\end{cases}\,,
\end{align}
where $z^* = 1/2( -1 + \sqrt{1+4a})$ is the stable fixed point of the dynamics \eqref{eq:eomZ} for $a > -1/4$. The drastic change of $\varrho_s(Z=z|A_n=a)$ at $a=-1/4$ can be understood by noting that Eq.~\eqref{eq:eomZfinal} undergoes a saddle-node bifurcation at that point, generating two fixed points of $Z$ for $A_n>-1/4$. For long times, $Z$ ends up at the stable of the two fixed points, $z^*$, in this case. For $A_n<-1/4$, on the other hand, there are no fixed points and $Z$ reaches $-\infty$ in a finite time. Because $Z=V/X$, this case corresponds to two particles overtaking each other, forming a caustic \cite{Gus16,Fal02,Wil06,Wil05}. Upon reaching $-\infty$, $Z$ is immediately reinjected at $+\infty$ to ensure $Z=V/X$ also after the caustic event \cite{Gus11b,Gus14c}. We now calculate the steady-state probability density $P_s(Z=z)$, using
\begin{align}
	P_s(Z=z) 	=& \int_{-\infty}^\infty \varrho_s(Z=z|A_n=a) P(A_n = a) \ed a \,,		\label{eq:zdist}
\end{align}
and substituting the expressions in Eqs.~\eqref{eq:kacrice} and \eqref{eq:zconddist} for $P(A_n=a)$ and $\varrho_s(Z=z|A_n=a)$, respectively. We obtain
\begin{align}
	P_s&(Z=z) =	 -\frac{2z(z+1/2)(z+1)}{\kappa^2}\ee^{-z^2(z+1)^2/(2\kappa^2)}\mathbf{1}_{[-1/2,0]}	\nn\\
			&+\int_{0}^{\infty}\frac{\sqrt{a}(a+1/4)}{\pi \kappa^2[(z+1/2)^2+a]}  \ee^{-(a+1/4)^2/(2\kappa^2)} \ed a \label{eq:pz}\,,
\end{align}
where, $\mathbf{1}_{S}$ denotes the indicator function of the set $S$. The steady-state distribution is the sum of two parts, corresponding to the two different solutions of $\varrho_s(Z=z|A_n=a)$. The first term in Eq.~\eqref{eq:pz} originates from the gradient range $A_n\geq-1/4$, that does not admit caustics, while the second one comes from the gradient range where caustics form. The result for $P_s(Z=z)$, Eq.~\eqref{eq:pz}, as a function of $z$ for $\kappa = 1/4$ is shown as the solid line in Fig.~\ref{fig:zdistr}. The distribution exhibits a peak at $z=-1/2$ and a sharp kink at $z=0$. For large $z$ the distribution has power-law tails $\propto \mathcal{J}/z^2$, shown as the dotted lines. The steady-state distribution for $z$ in the persistent-flow model is thus quite similar to the corresponding quantity in the white-noise model, shown as the dash-dotted line in Fig.~\ref{fig:zdistr}. We observe a similar peak at small $z$ and power-law tails with $z^{-2}$ scaling. 

The apparent similarity of the two distributions in these very different limits is due to the similar qualitative dynamics of $Z$. In the persistent-flow model, $Z$ equilibrates at $z^*$, close to $-1/2$, for the majority of realisations, $A_n\geq-1/4$, hence the peak at $z=-1/2$. For $A_n<-1/4$, $Z$ repeatedly approaches $-\infty$ and returns, leading to the power law tails $\sim z^{-2}$ in the distribution. In the white-noise model, on the other hand, the noisy dynamics of $Z$ spends most of its \textit{time} close to the stable fixed point at $z=0$. Occasionally, however, the gradient history allows $Z$ to pass the unstable fixed point at $z=-1$ and escape to infinity. The $z^{-2}$-tails of the distribution originate from caustic events $Z\to-\infty$. The large-$Z$ dynamics is universal and independent of the gradient $A(\bar x,t)$, because the deterministic $Z^2$-term in Eq.~\eqref{eq:eomZ} dominates for large enough $Z$.
\subsection{Lyapunov exponent}
\begin{figure}[t]
\includegraphics{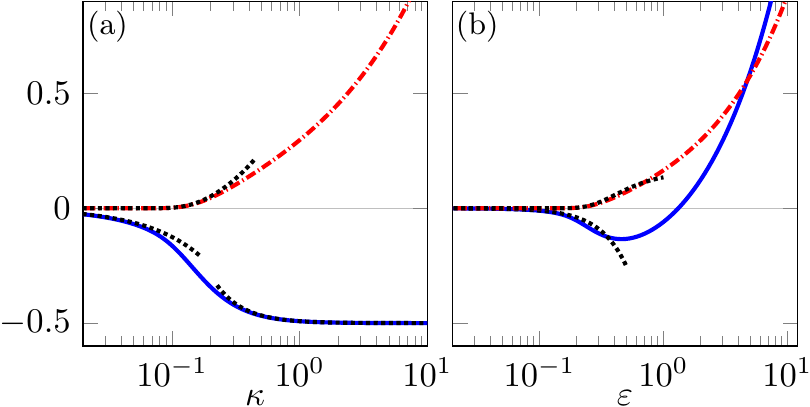}
\caption{Spatial Lyapunov exponent $\lambda$ and rate of caustic formation $\mathcal{J}$ in the different limits. (a) Persistent-flow limit results for $\lambda$, Eq.~\eqref{eq:lyap2}, (solid line) and $\mathcal{J}$, Eq.~\eqref{eq:caust2} (dash-dotted line) plotted as functions of $\kappa$. The dotted lines show the asymptotic behaviours for small and large values of $\kappa$ (see main text). (b) Corresponding 
results for the white-noise limit:  $\lambda$ (solid line) and $\mathcal{J}$ (dash-dotted line) plotted as functions of $\varepsilon$. The dotted lines show the asymptotic behaviours for small values of $\varepsilon$ (see main text).}\label{fig:lyapcaust}
\end{figure}
\begin{figure*}[t]
\includegraphics{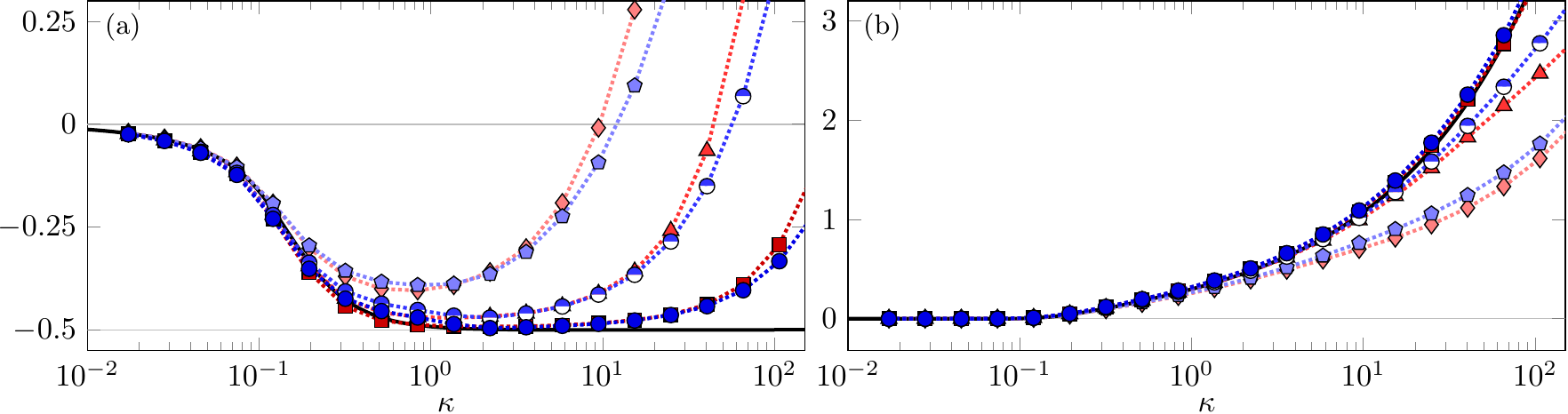}
\caption{Spatial Lyapunov exponents and rate of caustic formation as functions of $\kappa$ for different
values of $\Ku$. (a) Results for $\lambda$ obtained from direct numerical simulations of Eq.~\eqref{eq:lyapnum} plotted as pentagons ($\Ku = 10^1$), half-circles ($\Ku = 10^2$) and filled circles ($\Ku=10^3$). The corresponding results for $\lambda_A$ (Section \ref{sec:nsdpf}) are shown as diamonds ($\Ku=10^1$), triangles ($\Ku=10^2$) and squares ($\Ku=10^3$). The solid line is the persistent-flow model result, Eq.~\eqref{eq:lyap2}. (b): Results for $\mathcal{J}$ obtained from direct numerical simulations of Eq.~\eqref{eq:caustnum} plotted as pentagons ($\Ku = 10^1$), half-circles  ($\Ku = 10^2$) and filled circles ($\Ku=10^3$). The corresponding results for $\mathcal{J}_A$ (Section \ref{sec:nsdpf}) are shown as diamonds ($\Ku=10^1$), triangles ($\Ku=10^2$) and squares ($\Ku=10^3$). The solid line shows the persistent-flow model result, Eq.~\eqref{eq:caust2}.}\label{fig:lyapcaustdist}
\end{figure*}
The spatial Lyapunov exponent 
\begin{align}
	\lambda = \lim_{t\to\infty}\lim_{X(0)\to0}\frac1t\log \frac{|X(t)|}{|X(0)|}\,,
\end{align}
determines the long-time fate of particle separations. It is given by the steady-state expectation value $\lambda=\langle Z\rangle_s$ \cite{Gus16}:
\begin{align}
	\lambda 	&= \mathcal{P}\!\!\int_{-\infty}^{\infty} \!\!\!\!\ed z\,z P_s(Z=z)	= -\int_{-1/2}^0\!\!\!\!\ed z\, \ee^{-z^2(z+1)^2/(2\kappa^2)}\,,	\label{eq:lyap}
\end{align}
where $\mathcal{P}$ denotes the principle value of the integral. The second equality follows using Eq.~\eqref{eq:pz} and performing integrations by part. The integral on the right-hand side of Eq.~\eqref{eq:lyap} is always negative and can be expressed in terms of special functions as
\begin{align}\label{eq:lyap2}
	\lambda = -\frac{1}{2} + \frac{\, _2F_2\left(1,\frac{3}{2};\frac{7}{4},\frac{9}{4};-\frac{1}{32 \kappa ^2}\right)}{120 \kappa ^2}\,,
\end{align}
where $_2F_2$ denotes the generalised hypergeometric function. The solid line in Fig.~\ref{fig:lyapcaust}(a) shows the spatial Lyapunov exponent as a function of $\kappa$. For small $\kappa$, $\lambda$ scales as $\lambda\sim\langle A_n\rangle\propto -\kappa$, signalling clustering due to preferential concentration. For large $\kappa$, the spatial Lyapunov exponent asymptotes to $\lambda\sim-1/2+1/(120 \kappa^2)$ and thus approaches $-1/2$.  The fact that $\lambda$ is negative for all $\kappa$ indicates that the particles in the trapping regions move closer and closer towards each other and the particle dynamics eventually contracts to a point for all $\kappa$.

As a comparison, the Lyapunov exponent for the white-noise model is shown as the solid line in Fig.~\ref{fig:lyapcaust}(b). In the white-noise model, $\lambda$ is negative for small $\varepsilon$ and then changes sign and becomes positive at $\varepsilon_\text{c} \approx 1.33$. This behaviour is known as \lq{}path-coalescence transition\rq{} \cite{Wil03} or \lq{}aggregation-disorder transition\rq{} \cite{Deu85} and describes the transition to chaotic behaviour of the model for $\varepsilon>\varepsilon_\text{c}$. For small $\varepsilon$, $\lambda$ scales as $\lambda\sim-\varepsilon^2$, meaning that $\lambda \propto \St$ for small inertial parameter in both models. The scaling as a function of $\Ku$ is, however, different, as we observe $\lambda\propto \Ku$ in the persistent-flow model and $\lambda\propto\Ku^2$ in the white-noise model at small inertia parameter.
\subsection{Rate of caustic formation}\label{sec:caust}
The rate of caustic formation $\mathcal{J}$ is the rate at which $|X(t)|\to0$ while $|V(t)|>0$. That is, it is the rate at which particle trajectories cross at finite relative velocity. Globally speaking, it is the rate at which the phase-space manifold of the particles folds over. The rate of caustic formation is of particular importance for the calculation of collision rates between particles, where it leads to corrections to the Saffman-Turner formula for advected particles \cite{Saf55,Pum16,Fal07c,Vos13}. We compute $\mathcal{J}$ from the magnitude of the tail of the steady-state distribution $P_s(Z=z)$, which behaves as $P_s(Z=z)\sim\mathcal{J}/z^2$ for large $z$:
\begin{align}
	\mathcal{J} 	= \lim_{z\to\infty} z^2 P_s(Z=z)						=\!\! \int_{0}^{\infty} \!\!\!\!\ed a\frac{\sqrt{a}(a+1/4)}{\pi \kappa^2}  \ee^{-(a+1/4)^2/(2\kappa^2)}		\label{eq:caust}\,.
\end{align}
Again, the integral can be expressed in closed form using special functions. We obtain
\begin{align}\label{eq:caust2}
	\mathcal{J} = (2^\frac52 \pi)^{-1}\ee^{-1/(64\kappa^2)} K_{-\frac14}\left(\frac1{64\kappa^2}\right)\,,
\end{align}
where $K_\nu$ is the modified Bessel function of the second kind. In Fig.~\ref{fig:lyapcaust}(a) $\mathcal{J}$ is shown as the dash-dotted line. For small $\kappa$, $\mathcal{J}$ exhibits an exponentially small activation $\propto \ee^{-1/(32\kappa^2)}$. Exponential activations are common for the rate of caustic formation of inertial particles in turbulence. They have been observed both in statistical models  \cite{Gus16,Wil05,Wil06} and in numerical simulations \cite{Fal07c,Vos13}. However, the exponent of the activation differs, depending on the model and on the limit of $\Ku$ and $\St$ that is taken \cite{Gus16,Meh04,Gus13a,Mei17}.

That said, the corresponding result for the white-noise model is shown as the dash-dotted line in Fig.~\ref{fig:lyapcaust}(b). For small $\varepsilon$, $\mathcal{J}$ exhibits an activated behaviour $\propto \ee^{-1/(6\varepsilon^2)}$. In terms of $\Ku$ and $\St$, the activation has the exponents $-1/(96\,\Ku^2\St^2)$ and $-1/(18\,\Ku^2 \St)$ for the persistent-flow model and the white-noise model, respectively.
\subsection{Numerical simulations and deviations from persistent-flow limit}
\label{sec:nsdpf}
In order to study the validity of our analytical results for finite $\Ku$ and $\St$, we perform direct numerical simulations of the statistical model. To this end, we generate a one-dimensional Gaussian random field with correlation \eqref{eq:corr} and let a large number of particles evolve according to the dynamics \eqref{eq:eom}. The simulations are performed at large $\Ku$ ($10^1,10^2$ and $10^3$), and different values of $\St$, resulting in a range of values of the inertia parameter $\kappa = \sqrt{3}\, \Ku\St$.

The spatial Lyapunov exponent is obtained from direct numerical simulations by computing the long-time empirical mean of $Z$ along a particle trajectory $x(t)$ \cite{Gus16}:
\begin{align}\label{eq:lyapnum}
	\lambda = \lim_{t\to\infty} \frac1t \int_0^t Z(s) \ed s\,.
\end{align}
The rate of caustic formation is, in turn, calculated by counting the number $N(t)$ of the events $Z\to-\infty$ after a long time $t$, divided by $t$:
\begin{align}\label{eq:caustnum}
	\mathcal{J} = \lim_{t\to\infty} \frac{N(t)}{t}\,.
\end{align}
Fig.~\ref{fig:lyapcaustdist}(a) shows $\lambda$ obtained from Eq.~\eqref{eq:lyapnum} for $\Ku=10^1$, $10^2$ and $10^3$ as pentagons, half circles and filled circles respectively.  The solid line shows the analytic result \eqref{eq:lyap2}. For small $\kappa$, the results of direct numerical simulations and the theory agree well for all measured $\Ku$. This is expected, because for $\kappa\ll1$ and fixed large $\Ku$, the Stokes number is small, $\St\ll1$, and we are close to the persistent-flow limit. As $\Ku$ increases, the regime of validity for $\kappa$ increases and the theoretical result is approached. The persistent-flow theory says that $\lambda<0$ for all values of $\kappa$. 
We observe, however, that this prediction fails for finite (but large) values of $\Ku$. Instead, $\lambda$ obtained from the direct numerical simulations using Eq.~\eqref{eq:lyapnum} exhibits a path-coalescence transition, just as in the white-noise model, for some $\kappa_\text{c}$. Hence, the dynamics becomes chaotic and the particles cluster on a fractal set for $\kappa>\kappa_\text{c}$, instead of contracting to a point.

Fig.~\ref{fig:lyapcaustdist}(b) shows $\mathcal{J}$ obtained from direct numerical simulations of Eq.~\eqref{eq:caustnum} for $\Ku=10^1$, $10^2$ and $10^3$ as pentagons, half-circles and filled circles, respectively. The analytic result \eqref{eq:caust2} is, as before, shown as the solid line. We observe similar behaviour as for the spatial Lyapunov exponent $\lambda$: The results of direct numerical simulations agree with the theory for all $\Ku$ if $\kappa\ll1$ but clearly show deviations for larger values of $\kappa$. For increasing $\Ku$, the theoretical formula \eqref{eq:caust2} is approached. 

In conclusion, the analytical formulas for $\lambda$ and $\mathcal{J}$, Eqs.~\eqref{eq:lyap2} and \eqref{eq:caust2}, agree excellently with the results direct numerical simulations performed at finite values of $\Ku$ and $\St$, if the inertia parameter $\kappa$ is small.

\begin{figure}[t]
\includegraphics{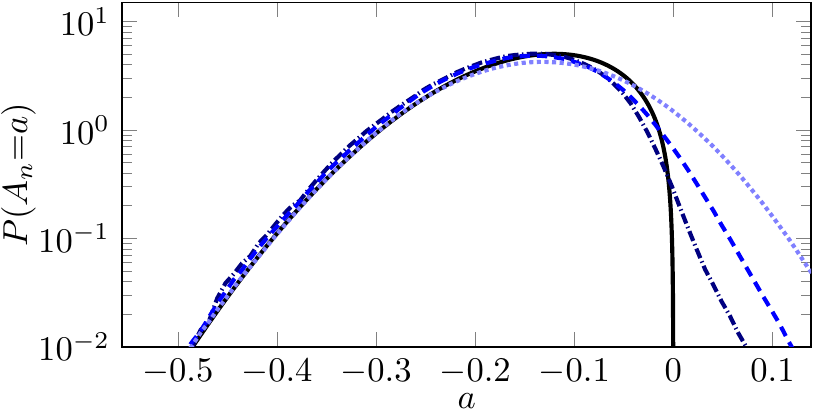}
\caption{Probability density $P(A_n=a)$ of fluid-velocity gradients for $\kappa = 0.12$ and different $\Ku$. Shown are results for $\Ku=10^1$ (dotted line), $\Ku=10^2$ (dashed line) and $\Ku = 10^3$ (dash-dotted line). The solid line corresponds to the distribution \eqref{eq:kacrice} that is approached in the limit $\Ku\to\infty$.}\label{fig:Adistr}
\end{figure}

Finally we discuss the origin of the deviations between the analytical formulae [Eqs.~\eqref{eq:lyap2} and \eqref{eq:caust2}] and our numerical results when $\Ku$ is large but finite, and $\kappa$ is not small. These deviations originate from two different effects: First, the migration time between trapping regions is not negligible in general -- which changes
the gradient distribution so that Eq.~(\ref{eq:kacrice}) fails. Second, outside the persistent-flow limit the fluid velocity gradient changes during the relaxation of the particle velocity, so that Eq.~(\ref{eq:zconddist}) breaks down.

Our theory uses the gradient distribution $P(A_n=a)$ as an input into Eqs.~\eqref{eq:zdist} so that the importance of the first effect can be tested by modifying $P(A_n=a)$ appropriately. While  $P(A_n=a)$ is given by Eq.~\eqref{eq:kacrice} in the persistent-flow limit, it is generally determined by a complicated interplay between the dynamics of the inertial particles and the fluid motion. For that reason, so we do not have a closed-form expression for $P(A_n=a)$ for finite $\Ku$ and for values of $\kappa$ that are not small. Instead, we compute $P(A_n=a)$ from our direct numerical simulations by measuring $A_n$ along particle trajectories. Using the numerically computed gradient distribution as an input into Eqs.~\eqref{eq:zdist}, we then compute the Lyapunov exponent and the rate of caustic formation with the help of the first expressions in Eqs.~\eqref{eq:lyap} and \eqref{eq:caust}. We denote the Lyapunov exponent computed in this way by $\lambda_A$, and the rate of caustic formation by $\mathcal{J}_A$.
Note that these quantities are still calculated by keeping the gradients constant on the time scales of the particle dynamics, only the gradient distribution is adjusted. Comparing $\lambda_A$ and $\mathcal{J}_A$ to $\lambda$ and $\mathcal{J}$ obtained from our direct numerical simulations thus allows us to quantify the relevance of the finite migration time between trapping regions.
 
Our simulation results for $P(A_n=a)$ are shown in Fig.~\ref{fig:Adistr}. We observe convergence towards Eq.~\eqref{eq:kacrice} for large $\Ku$, but positive gradients have non-zero probability for all finite $\Ku$. The probability for positive gradients decreases rapidly, however, as the persistent-flow limit is approached. For $\kappa = 0.12$ and $\Ku = 10^3$, for example, the probability of observing positive gradients is very small, $P(A_n>0) \approx 4\cdot10^{-3}$.
 
 The results for $\lambda_A$ and $\mathcal{J}_A$ are shown as the diamonds, triangles and squares in Fig.~\ref{fig:lyapcaustdist}(a) and (b), respectively (see Figure caption for details). We observe that $\lambda_A$ and $\mathcal{J}_A$ agree qualitatively with the results for $\lambda$ and $\mathcal{J}$ from our direct numerical simulations.  Surprisingly, the path-coalescence transition for $\lambda$ is reproduced quite accurately by $\lambda_A$. This means that deviations from the persistent-flow limit seen in Fig.~\ref{fig:lyapcaust} at large but finite $\Ku$ and small but finite $\kappa$ are caused predominantly by the fact
 that the distribution $P(A_n = a)$ of gradients  changes because the migration between traps must be taken into account. Fig.~\ref{fig:lyapcaustdist} shows that the persistent approximation works very well for our model, deviations occur only at large values of $\kappa$. This is in contrast to white-noise models where the Lyapunov exponent and the rate of caustic formation are strongly influenced by the fluctuations of the fluid-velocity gradients experienced in the past \cite{Gus16}.
\section{Conclusions and discussion}\label{sec:conc}
We have analysed a one-dimensional, statistical model for heavy particles in turbulence in the limit of large flow persistence and weak particle inertia. Because the one-dimensional fluid-velocity field is compressible, it exhibits long-lived trapping regions where particles tend to accumulate. As the particles spend most of their time close to these traps, we have shown that the fluid-velocity gradient statistics in these regions determine the spatial Lyapunov exponent $\lambda$ and rate of caustic formation $\mathcal{J}$. We find that the rate of caustic formation is an increasing function of the inertia parameter $\kappa = \sqrt{3}\Ku\,\St$ in the persistent-flow limit, with exponential activation. The spatial Lyapunov exponent, by contrast, decreases monotonously as $\kappa$ increases and remains always negative. This indicates strong spatial clustering of the particles on point-like sets. 

Our analytical results for $\lambda$ and $\mathcal{J}$ agree with those of direct numerical simulations of our model at large $\Ku$ as long as $\kappa$ is small. For larger values of $\kappa$ we observe deviations from the analytical results derived in the persistent-flow limit. Our analysis shows that these deviations result from the fact that migration between traps must be taken into account. This changes the positive-gradient tail of the distribution of particle-velocity gradients. We have not yet found a reliable analytical approximation for these tails.

The persistent approximation remains quite accurate even for large values of the inertia parameter $\kappa$. This means that the Lyapunov exponent and the rate of caustic formation are essentially determined by instantaneous, local properties of the flow. This is in contrast to the white-noise limit, where the history of fluid-velocity gradients experienced in the past matters \cite{Gus16}.

The existence of the long-lived trapping regions in the flow is an important ingredient for the one-dimensional persistent-flow model. In statistically homogeneous and isotropic flows such trapping regions exist only if the flow is compressible. Therefore, we expect particles in higher dimensional flows to exhibit a similar behaviour to that described here only if the flow is persistent, sufficiently {\em compressible}, and statistically homogeneous and isotropic.

Let us briefly discuss the relevance of the present results for incompressible, higher-dimensional flows, and for incompressible, fully developed turbulence in particular. 
As mentioned in Section \ref{sec:prob}, the Kolmogorov time $\tauk$ is analogous to the advection time scale $\taua$ in our model. However, in incompressible turbulence, and in incompressible flows in general, there are no trapping regions where advected particles stay for a long time. This means that both advected and heavy particles typically travel long distances, experiencing different fluid-velocity gradients along their way. Hence,
the persistent-flow model does not directly apply to heavy particles in incompressible turbulence.

There might, however, be certain situations in turbulent flows where the combined dynamics of the fluid and the particles leads to trapping of particle trajectories. In these situations it may still be a good approximation to treat the fluid-velocity gradients at the particle position as persistent compared to the particle dynamics. Examples could be bubbles trapped in long-lived turbulent vortex-tubes by Tchen's force \cite{Che47}, or heavy particles in coherent turbulent structures with large fluid-velocity gradients, but small mean flow.
\section{Acknowledgements}
We thank Kristian Gustavsson for discussions and for his comments on the manuscript. This work was supported by Vetenskapsr\aa{}det (Grant No. 2017-3865), Formas (Grant No. 2014-585), and by the grant \lq{}Bottlenecks for particle growth in turbulent aerosols\rq{} from the Knut and Alice Wallenberg Foundation, Dnr. KAW 2014.0048. The numerical computations were obtained using resources provided by C3SE and SNIC.
\appendix
\section{Calculation of observables from Eqs.\eqref{eq:eomXVfinal}}\label{sec:obsdirect}
Instead of solving the dynamics for $Z$ as done in the main text, we show here alternative derivations for Eqs.~\eqref{eq:lyap2} and \eqref{eq:caust2} using the dynamics for $X$ instead. To this end, we first write Eqs.~\eqref{eq:eomXVfinal} as
\begin{align}
	\ddot X +\dot X -  A_n X = 0\,.
\end{align}
Because $A_n$ is constant, this equation can be readily solved for $X$. We obtain
\begin{align}\label{eq:Xsol}
	X(t) = C_1 \ee^{\lambda_+ t} + C_2 \ee^{\lambda_-t}	\,,
\end{align}
where $\lambda_\pm = -1/2 \pm \sqrt{1/4+A_n}$ and $C_1$ and $C_2$ are constants that depend on the initial conditions. Because $\Re[\lambda_+] \geq \Re[\lambda_-]$ for all realisations $A_n$, either the first term in Eq.~\eqref{eq:Xsol} dominates, or the two terms are of the same order. We may therefore assume for simplicity that $C_2 = 0$. From the remaining exponent $\lambda_+$ we can now compute both the Lyapunov exponent $\lambda$ and the rate of caustic formation $\mathcal{J}$. The former is obtained by averaging the real part of $\lambda_+$, $\Re[\lambda_+]$, over the realisations of $A_n$. We have
\begin{align}\label{eq:lyapdirect}
	\lambda 	&= \int_{-\infty}^\infty \!\!\!\!\ed a\, \Re\left[\lambda_+\big|_{A_n=a}\right] P(A_n = a)\,,	\nn\\
			&= -\frac12 + \int_{-1/4}^{\infty}\!\!\!\!\ed a\,\sqrt{1/4 + a} P(A_n = a)\,.
\end{align}
This expression is equal to Eq.~\eqref{eq:lyap}, if we substitute the gradient distribution \eqref{eq:kacrice} for $P(A_n=a)$ and make the change of variables $a\to z(a) = z^* = \lambda_+\big|_{A_n = a}$.

The rate of caustic formation $\mathcal{J}$ is obtained by the rate at which $|X|\to0$ while $|\dot X|>0$, averaged over the realisations of $A_n$. For fixed $A_n$, $X$ goes to zero at rate $\Im [\lambda_+]/\pi$, because $X$ passes zero twice in time intervals of length $2\pi/\Im[\lambda_+]$. Thus, we obtain for $\mathcal{J}$,
\begin{align}\label{eq:caustdirect}
	\mathcal{J}	&= \frac1\pi \int_{-\infty}^\infty\!\!\!\!\ed a\, \Im \left[\lambda_+\big|_{A_n = a}\right] P(A_n=a)	\nn	\\
				&= \frac1\pi \int_{-\infty}^{-1/4}\!\!\!\!\ed a\, \sqrt{-1/4-a}  P(A_n=a)\,,	
\end{align}
which, again, is equal to Eq.~\eqref{eq:caust} after substituting the gradient distribution \eqref{eq:kacrice} for $P(A_n=a)$, changing variables to $a\to -a-1/4$, and performing an integration by parts.
\hfill\eject
\end{document}